\documentclass[a4paper]{spie}  

\usepackage{amsmath,amsfonts,amssymb}
\usepackage{graphicx}
\usepackage[exponent-product=\cdot,separate-uncertainty=true]{siunitx}
\usepackage[colorlinks=true, allcolors=blue]{hyperref}
\newcommand{\e}{\ensuremath{\text{e}^\text{-}}}

\title{Spectroscopic performance of flight-like\\DEPFET sensors for Athena's WFI}

\author[a]{J.~Müller-Seidlitz}
\author[a]{R.~Andritschke}
\author[a]{M.~Bonholzer}
\author[a]{V.~Emberger}
\author[a]{G.~Hauser}
\author[a]{M.~Herrmann}
\author[b]{P.~Lechner}
\author[a]{A.~Mayr}
\author[a,$\dagger$]{J.~Oser}
\affil[a]{Max-Planck-Institute for Extraterrestrial Physics, Gießenbachstraße~1,~85748~Garching~bei~München,~Germany}
\affil[b]{Semiconductor Laboratory of the Max-Planck-Society, Otto-Hahn-Ring~6,~81739~München,~Germany}

\authorinfo{${}^\dagger$ now at Thorlabs GmbH, Münchner Weg 1, 85232 Bergkirchen, Germany\\Corresponding author:\\\hspace*{2em}J. Müller-Seidlitz: E-mail: johannes.mueller-seidlitz@mpe.mpg.de, Telephone: +49 (0)89 30 000 3509}

\begin{document} 
\maketitle

\begin{abstract}
The Wide Field Imager for the Athena X-ray telescope is composed of two back side illuminated detectors using DEPFET sensors operated in rolling shutter readout mode: A large detector array featuring four sensors with 512×512 pixels each and a small detector that facilitates the high count rate capability of the WFI for the investigation of bright, point-like sources. Both sensors were fabricated in full size featuring the pixel layout, fabrication technology and readout mode chosen in a preceding prototyping phase. We present the spectroscopic performance of these flight-like detectors for different photon energies in the relevant part of the targeted energy range from \SI{0.2}{\keV} to \SI{15}{\keV} with respect to the timing requirements of the instrument. For \SI{5.9}{\keV} photons generated by an \textsuperscript{55}Fe source the spectral performance expressed as Full Width at Half Maximum of the emission peak in the spectrum is \SI{126.0}{\eV} for the Large Detector and \SI{129.1}{\eV} for the Fast Detector. A preliminary analysis of the camera's signal chain also allows for a first prediction of the performance in space at the end of the nominal operation phase.
\end{abstract}


\section{INTRODUCTION}
\label{sec:intro}  

The DEPFET (DEpleted P-channel Field-Effect Transistor) \cite{kemmer87a} is the chosen X-ray detection and first signal amplification principle for the MPE-led development of the Wide Field Imager (WFI) of Athena, a large class mission of ESA’s Cosmic Vision program.\cite{nandra13,rau16}

The WFI consists of two units.\cite{meidinger20} The Large Detector Array composed of four 512×512 pixel matrices and a field of view of \SI{40}{\arcminute}×\SI{40}{\arcminute} has a timing requirement of \SI{\le 5}{\ms}, which demands a readout time of \SI{\le 9.8}{\us} per sensor row for the applied rolling shutter readout mode. For a second, smaller sensor with 64×64 pixels for the observation of bright, point-like sources, a row-wise readout speed of \SI{2.5}{\us} as well as a two-row parallel readout is needed to fulfill the pile-up and throughput requirements.

\begin{figure}[ht]
	\begin{center}
		\begin{tabular}{c} 
			\includegraphics[height=10cm]{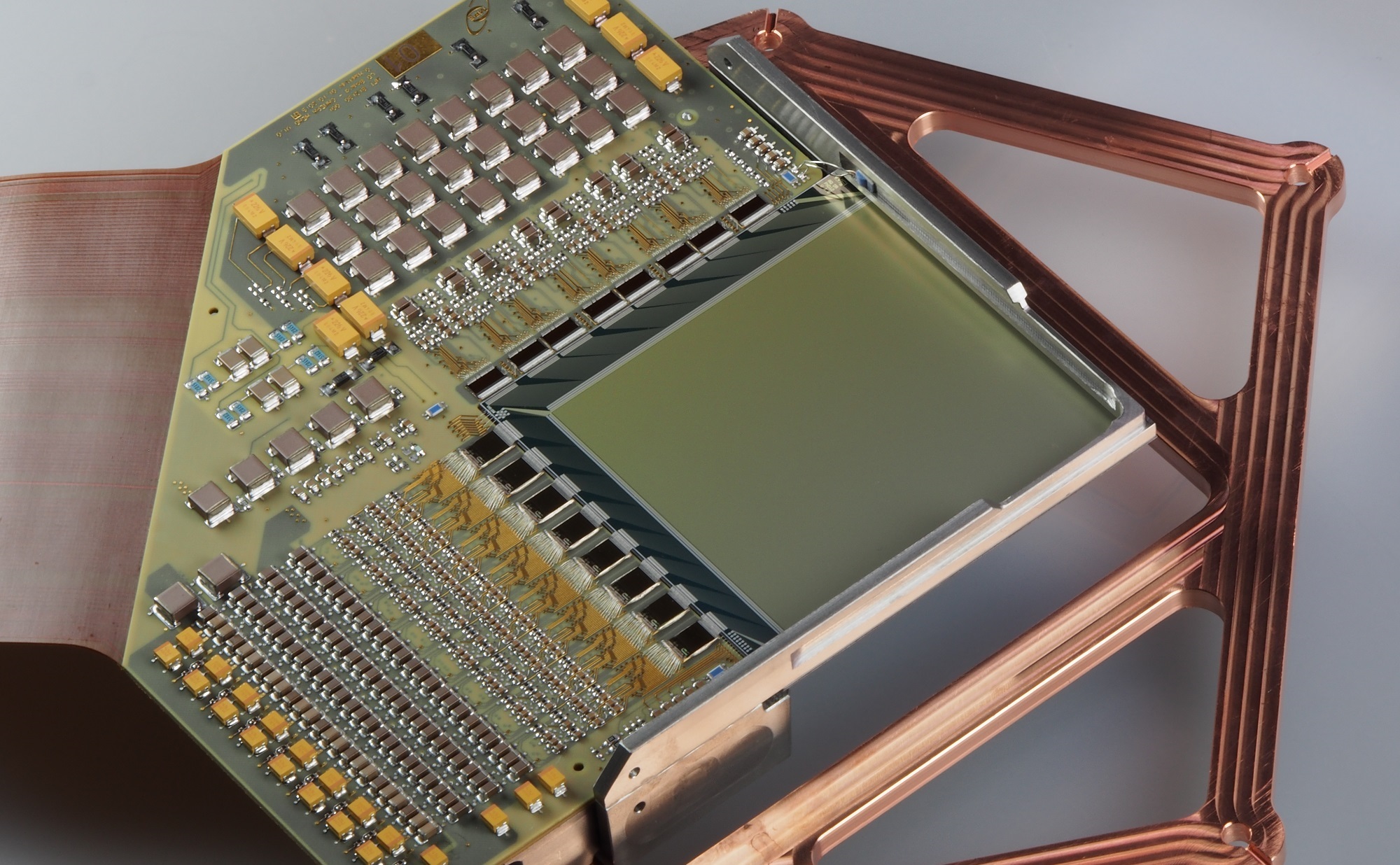}
		\end{tabular}
	\end{center}
	\caption[LD]
		{ \label{fig:ld-module} 
		A Large Detector module consisting of 512×512 DEPFET pixels. It is steered row-wise by eight Switcher ASICs\cite{fischer03} (top left) and read out column-parallel via eight Veritas readout ASICs\cite{porro14,herrmann18} (bottom left). The copper frame on the right is used for the mounting in the measurement vacuum chamber. Four of such Large Detectors (quadrants) will build the Large Detector Array of Athena's Wide Field Imager.}
\end{figure} 

After the pixel layout, the fabrication technology and the readout mode for the detector were fixed in a preceding prototyping phase,\cite{treberspurg16,treberspurg17,treberspurg18} DEPFET arrays with the full flight size were fabricated. A fully assembled Large Detector is shown in \autoref{fig:ld-module}. The pixel size of both detectors is \SI{130}{\um}×\SI{130}{\um}.

In a DEPFET active pixel sensor for X-ray astrophysics, the electrons generated in the sensitive volume of the back side illuminated device are collected under the DEPFET channel and influence its conductivity by inducing mirror charges into the transistor channel. It enables the measurement of their number that corresponds to the energy of the charge generating incident photon. Quasi Fano-limited spectral performance in combination with a high readout speed can be achieved. To realize a good quantum efficiency even at \SI{15}{\keV}, the sensitive volume is maximized by a full depletion of the silicon semiconductor over the entire sensor thickness of \SI{450}{\um}.

\section{Measurement and Analysis Method}
\label{sec:method}

\begin{figure}[ht]
	\begin{center}
		\begin{tabular}{c} 
			\includegraphics[height=5cm]{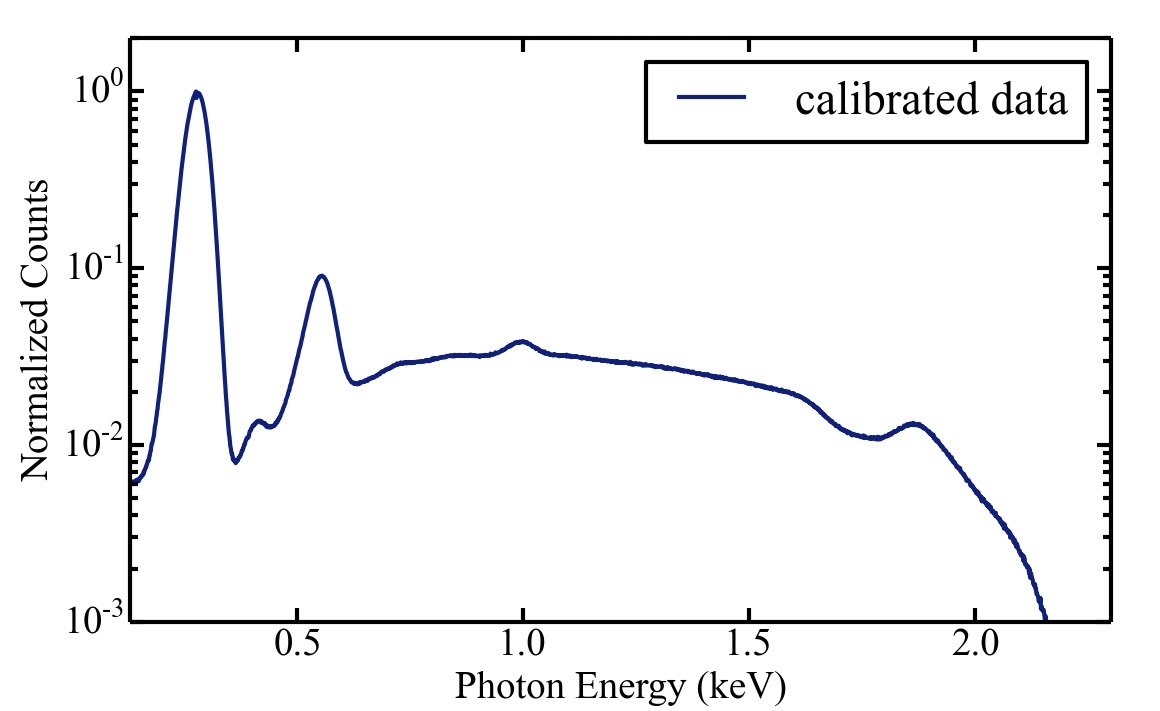}~~~
			\includegraphics[height=5cm]{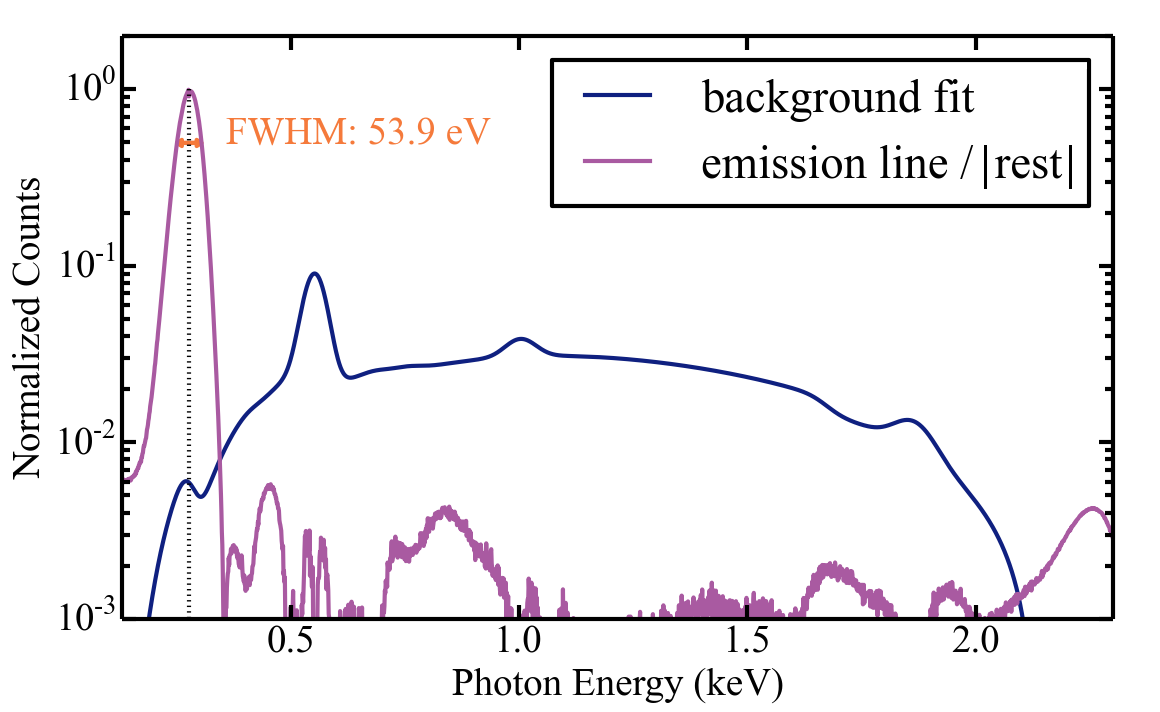}
		\end{tabular}
	\end{center}
	\caption[LD]
		{ \label{fig:fitexpl} 
		On the left, the measured spectrum of a carbon K$\alpha$ emission is shown. Due to charge losses, the gain for low energetic photons seems to decrease.\cite{muellerseidlitz18} A calibration to the emission line of interest shifts the more energetic photons to even higher energies while there is no energy dependence of the gain in the detector itself. Therefore, other emission lines---like O K$\alpha_{1,2}$---are shifted and also the continuum ends above \SI{2}{\keV} even though the electric field for the acceleration of electrons was limited to \SI{2}{\kV}. The fit accounts for these effects. On the right, the fit---excluding the emission line itself---as well as the residual is shown. It is the absolute value of the difference between the data and the fit. It contains the emission line with all effects that broaden it and the inaccuracies of the fit which are always below \SI{0.6}{\percent} of the main peak height.}
\end{figure}

Apart from the standard calibration source used for most of the measurements throughout the development phase---a radioactive \textsuperscript{55}Fe source---all the emission lines were produced with an X-ray tube available at the MPE laboratories. A filament is heated to emit electrons. They are accelerated and focused on a target of the desired material. Atoms in the target are ionized and emit characteristic radiation. The drawback is a contamination of the spectrum by bremsstrahlung emitted while the electrons are decelerated in the target material. The bremsstrahlung adds a continuous contribution up to the maximum energy $E_{max}$ that the electrons gained in the accelerating electric field. Its distribution is described by Kramers' law\cite{kramers23}
\begin{equation}
\label{eq:kramers}
\Psi(E) = \frac{K}{2 \pi c} \left(E_{max} - E\right)
\end{equation}
with $K$ proportional to the atomic number of the target element and the speed of light in vacuum $c$. To reduce the influence of the continuum on the spectrum, filters are placed into the ray path. In addition, the on-chip optical blocking filter, that reduces the optical loading on the detector during operation in space, affects the spectrum that is detected by the sensor. Using Kramers' law and filter transmission data from Henke et al.\cite{henke93}, a model for the continuum was generated. Known emission lines---apart from the one of interest---were added and an uncertainty according to Fano statistics\cite{fano47} was applied to model the measured spectrum. To obtain proper values for the spectral performance, the resulting fit of known components was subtracted from the measurement data. The remaining emission line including charge losses is then used to determine the Full Width at Half Maximum, FWHM. The widths of the manganese emission lines were determined without a background fit and its subtraction. In \autoref{fig:fitexpl}, an example of such a fit to measured data is given.

All measurements were performed at a temperature of \SI{-60}{\celsius} at the DEPFET sensor and at about \SI{0}{\celsius} at the front-end electronics which comprises the steering and readout ASICs (Application-Specific Integrated Circuit). To avoid icing and the absorption of the X-ray photons before they hit the sensor, the entire detector modules are operated in vacuum. All measurements were performed in continuous rolling shutter mode of the full sensor frame. The readout process and the multiplexing of the readout ASIC output data\cite{herrmann22} was always set to the slowest feasible speed to achieve the best performance possible at a given exposure time. The readout of the Fast Detector is split into two halves to gain a factor of two in speed.

For the spectra, event patterns of up to 2×2 pixels were considered. Larger patterns can only be generated by pile-up---also with a noise excess---or massive particles entering the sensor and are discarded. All spectroscopic performance results are given for all the accepted patterns. The performance of events that deposit all their charge carriers in a single pixel is typically a few electronvolts better. A limitation to those would result in a loss of a large fraction of detected photons and is not an option for a detector used in astrophysics.

\section{Results}
\label{sec:results}

The pre-flight production for Athena's WFI delivered DEPFET sensors of full size featuring the pixel layout, fabrication technology and readout mode designated for the flight modules. Using detailed inspection methods and repair effort on pixel level as well as improved fabrication steps, the yield was significantly increased and the overall homogeneity improved.\cite{bonholzer22} This resulted in the first functioning DEPFET sensors of this size. The obtained spectral performances of the two different detectors types designated for the WFI of Athena using the method described in \autoref{sec:method} are summarized in \autoref{tab:specperf}.

\begin{table}[ht]
\caption{The spectral performance of a Large (LD) and Fast Detector (FD) as determined from the measurements with different emission lines. The theoretical Fano limit is given for comparison. The Large Detector can be operated with a higher readout speed than required in exchange for a degradation of the performance. Missing values are due to failed measurements which need to be redone.} 
\label{tab:specperf}
\begin{center}       
\begin{tabular}{l|r|r|r|r|r|r}
\multicolumn{3}{c|}{~} & \multicolumn{3}{c|}{LD FWHM} & FD FWHM \\
Emission line & \multicolumn{1}{l|}{Energy} & Fano limit & $t_{exp} = \SI{5.00}{\ms}$ & $t_{exp} = \SI{2.00}{\ms}$ & $t_{exp} = \SI{1.28}{\ms}$ & $t_{exp} = \SI{80}{\us}$ \\
\hline
C K$\alpha_{1,2}$  & \SI{0.2770}{\keV}	& \SI{26}{\eV}	& \SI{53.9}{\eV}	& \SI{59.9}{\eV}	& \SI{65.4}{\eV}	& \SI{59.5}{\eV} \\
O K$\alpha_{1,2}$  & \SI{0.5249}{\keV}	& \SI{36}{\eV}	& \SI{56.0}{\eV}	& \SI{62.7}{\eV}	& \SI{67.5}{\eV}	& \SI{60.6}{\eV} \\
Zn L$\alpha_{1,2}$ & \SI{1.0117}{\keV}	& \SI{50}{\eV}	& \SI{61.1}{\eV}	& \SI{68.4}{\eV}	& \SI{74.9}{\eV}	& \SI{66.4}{\eV} \\
Al K$\alpha_1$     & \SI{1.4867}{\keV}	& \SI{60}{\eV}	& \SI{69.8}{\eV}	& \SI{75.3}{\eV}	& \SI{79.6}{\eV}	& \SI{74.7}{\eV} \\
Ag L$\alpha_1$     & \SI{2.9843}{\keV}	& \SI{85}{\eV}	& \SI{90.2}{\eV}	& \SI{94.2}{\eV}	& \SI{97.9}{\eV}	& \SI{93.3}{\eV} \\
Ti K$\alpha_1$     & \SI{4.5108}{\keV}	& \SI{105}{\eV}	& \SI{111.2}{\eV}	& \SI{117.2}{\eV}	& \SI{119.4}{\eV}	& \SI{113.8}{\eV} \\
Cr K$\alpha_1$     & \SI{5.4147}{\keV}	& \SI{115}{\eV}	& \SI{121.2}{\eV}	& \multicolumn{1}{c|}{---}	& \SI{128.6}{\eV}	& \SI{121.2}{\eV} \\
Mn K$\alpha_{1,2}$ & \SI{5.8951}{\keV}	& \SI{120}{\eV}	& \SI{126.0}{\eV}	& \SI{129.5}{\eV}	& \SI{132.2}{\eV}	& \SI{129.1}{\eV} \\
Fe K$\beta_{1,3}$  & \SI{7.0580}{\keV}	& \SI{131}{\eV}	& \SI{138.4}{\eV}	& \SI{140.2}{\eV}	& \SI{142.8}{\eV}	& \SI{139.4}{\eV} \\
\end{tabular}
\end{center}
\end{table}

\begin{figure}[ht]
	\begin{center}
		\begin{tabular}{c} 
			\includegraphics[height=5cm]{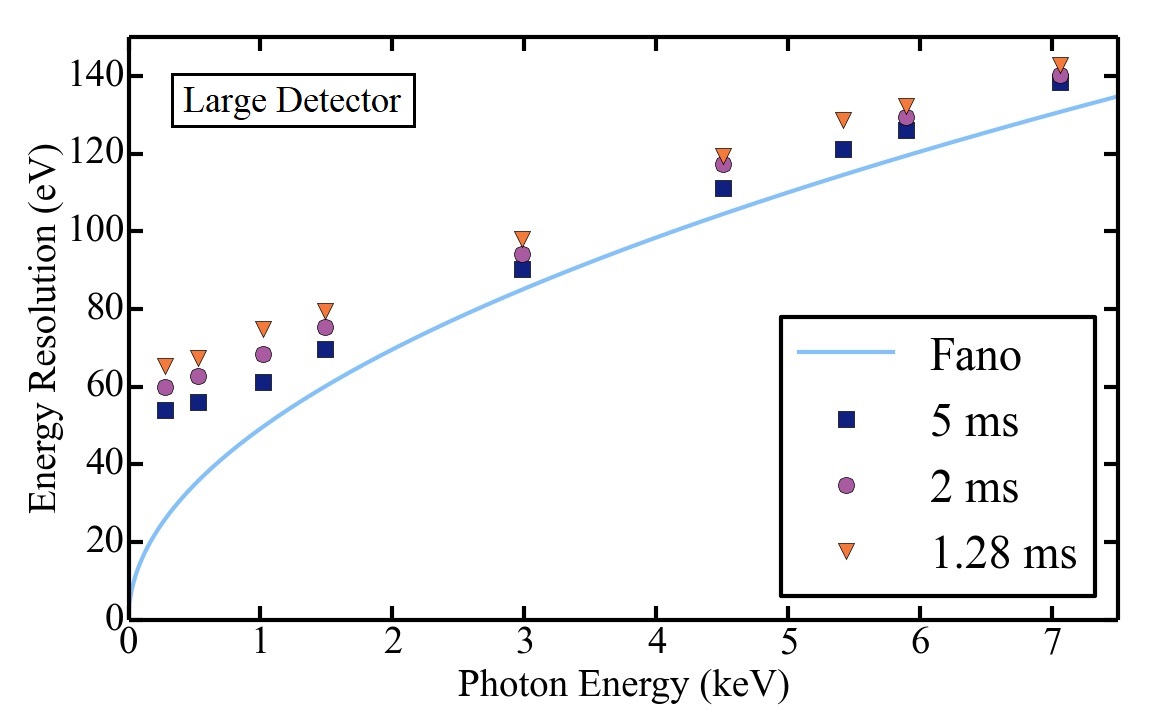}~~~
			\includegraphics[height=5cm]{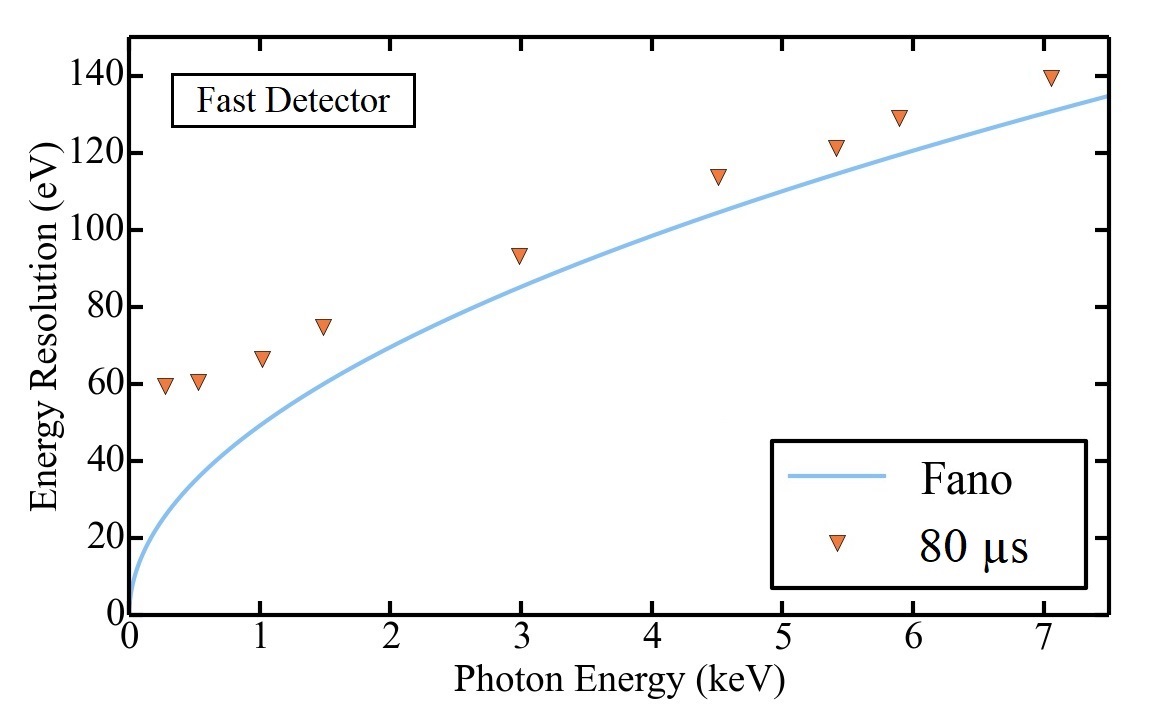}
		\end{tabular}
	\end{center}
	\caption[LD]
		{ \label{fig:specperf} 
		Visualized spectral performance of a Large (left) and a Fast Detector (right). The shown data are given in \autoref{tab:specperf}. The given times are the exposure times and therefore the time resolution of the sensor. For the readout, the full exposure time is used which allows for a better performance at a lower time resolution. The increasing difference between the theoretical Fano limit and the measured performance for lower photon energies is caused by charge losses at the entrance window. Low energetic photons have a lower penetration depth and the probability of charge losses into insensitive volume increases.}
\end{figure}

The Fano limited theoretical performance is calculated via the following expression.
\begin{equation}
\label{eq:fano}
\text{FWHM} = 2 \sqrt{2 \ln{2}} \sqrt{F \omega E}
\end{equation}
with the material dependent Fano factor $F = 0.118$ and the mean electron--hole pair creation energy $\omega = 3.7\,\e\text{/eV}$ for the operation temperature of \SI{-60}{\celsius} of the silicon sensor.\cite{lowe07}

The Large Detector suffers from a few noisy columns as well as rows with noisy pixels. In both cases, the origins are still unknown but are located in the DEPFET sensor. The additional noise in the columns can be eliminated with a modified common mode correction.

Even though the design of the readout chain is limited to \SI{2.5}{\us} per row, resulting in a minimum exposure time of \SI{1.28}{\ms} for the Large Detector, the flexible laboratory setup allows for higher speeds. For a frame time of \SI{1}{\ms}, a spectral performance of \SI{136.3}{\eV}~FWHM (\SI{129.5}{\eV} for single pixel events) at \SI{5.9}{\keV} photon energy was achieved.

\section{Performance Analysis}
\label{sec:analysis}

All results presented so far are obtained from laboratory measurements. Such measurements demonstrate the capabilities of the detectors under test but do not account for additional effects which might degrade the performance in a relevant environment. To be able to predict the spectroscopic potential, the entire signal chain and all aspects that might influence it were analyzed. For the degradation over time, first Total Ionizing Dose (TID) tests were performed.\cite{emberger22} The results of the first noise component analysis are summarized in \autoref{tab:perfana}. There are two types of contributions to the spectral performance: noise components $\sigma_i$ and further, non-noise effects $\Delta \text{FWHM}_i$ that broaden an emission line linearly.
\begin{equation}
\label{eq:fwhm}
\text{FWHM}(E) = 2 \sqrt{2 \ln{2}} \sqrt{F \omega E + \omega^2 \Sigma_i \sigma_i^2} + \Sigma_i \Delta \text{FWHM}_i(E)
\end{equation}
Most of the components listed in \autoref{tab:perfana} are already influencing the measurements taken in our laboratories. Only the degradation of shot and read noise, the photon background and potential electromagnetic (EM) emission from other parts of the Athena satellite will decrease the performance further. To get an impression of their impact, those components were added to the measured data from \autoref{tab:specperf}. The resulting performances are shown in \autoref{tab:specperfdeg}. Due to the quadratic addition of noise components, the effect on smaller numbers and, therefore, the performance at lower energies is slightly larger. While the degradation at low energies is about \SI{3}{\eV}, it is below \SI{2}{\eV} for higher energies. Another finding of the detailed analysis was, that the difference between the theoretical Fano limit and the measured data---which cannot be explained just by a line-broadening due to noise---is caused by threshold effects during the event recombination, which depends on the noise anyway.

\begin{table}[ht]
\caption{The individual components that contribute to the noise and, thereby, spectral performance of the two detectors. The analysis is done at \SI{1}{\keV} and \SI{7}{\keV} that are the energies at which performance requirements for the WFI instrument exist. They are \SI{\le 80}{\eV}~FWHM and \SI{\le 170}{\eV}~FWHM, respectively. The values represent the contribution to one event which typically spreads over multiple pixels. The factors multiplied to the noise of a single pixel are $1.26$ and $1.42$ for \SI{1}{\keV} and \SI{7}{\keV}, respectively.}
\label{tab:perfana}
\begin{center}       
\begin{tabular}{l|r|r|r|r}
 & \multicolumn{2}{c|}{LD} & \multicolumn{2}{c}{FD} \\
 & \multicolumn{1}{c|}{\SI{1}{\keV}} & \multicolumn{1}{c|}{\SI{7}{\keV}} & \multicolumn{1}{c|}{\SI{1}{\keV}} & \multicolumn{1}{c}{\SI{7}{\keV}} \\
\hline
Fano & \SI{21.5}{\eV}~RMS & \SI{56.0}{\eV}~RMS & \SI{21.5}{\eV}~RMS & \SI{56.0}{\eV}~RMS \\
Shot Noise & \SI{7.1}{\eV}~RMS & \SI{7.9}{\eV}~RMS & \SI{0.9}{\eV}~RMS & \SI{1.0}{\eV}~RMS \\
Photon Background & \SI{3.8}{\eV}~RMS & \SI{3.8}{\eV}~RMS & \SI{0.2}{\eV}~RMS & \SI{0.2}{\eV}~RMS \\
\hline
Power Supplies & \SI{2.0}{\eV}~RMS & \SI{2.0}{\eV}~RMS & \SI{2.0}{\eV}~RMS & \SI{2.0}{\eV}~RMS \\
\hline
Switcher & \SI{0.0}{\eV}~RMS & \SI{0.0}{\eV}~RMS & \SI{0.0}{\eV}~RMS & \SI{0.0}{\eV}~RMS \\
Read Noise & \SI{10.8}{\eV}~RMS & \SI{12.1}{\eV}~RMS & \SI{11.3}{\eV}~RMS & \SI{12.8}{\eV}~RMS \\
Veritas & \SI{7.1}{\eV}~RMS & \SI{8.0}{\eV}~RMS & \SI{5.7}{\eV}~RMS & \SI{6.4}{\eV}~RMS \\
Bandwidth Limits & \SI{0.2}{\eV}~RMS & \SI{0.2}{\eV}~RMS & \SI{0.2}{\eV}~RMS & \SI{0.2}{\eV}~RMS \\
Ext. EM Emission & \SI{0.1}{\eV}~RMS & \SI{0.1}{\eV}~RMS & \SI{0.1}{\eV}~RMS & \SI{0.1}{\eV}~RMS \\
\hline
ADC & \SI{3.0}{\eV}~RMS & \SI{3.0}{\eV}~RMS & \SI{3.0}{\eV}~RMS & \SI{3.0}{\eV}~RMS \\
OnBoard Pipeline & \SI{0.7}{\eV}~RMS & \SI{0.7}{\eV}~RMS & \SI{0.7}{\eV}~RMS & \SI{0.7}{\eV}~RMS \\
\hline
Ground Pipeline & \SI{1.5}{\eV}~RMS & \SI{6.0}{\eV}~RMS & \SI{1.5}{\eV}~RMS & \SI{6.0}{\eV}~RMS \\
\hline
Charge Losses & \SI{6.0}{\eV}~FWHM & \SI{6.0}{\eV}~FWHM & \SI{6.0}{\eV}~FWHM & \SI{6.0}{\eV}~FWHM \\
Non-Linearity & \SI{0.5}{\eV}~FWHM & \SI{0.5}{\eV}~FWHM & \SI{0.5}{\eV}~FWHM & \SI{0.5}{\eV}~FWHM \\
Energy Misfits & \SI{0.1}{\eV}~FWHM & \SI{0.1}{\eV}~FWHM & \SI{0.1}{\eV}~FWHM & \SI{0.1}{\eV}~FWHM \\
\end{tabular}
\end{center}
\end{table}

\begin{table}[ht]
\caption{Estimated mean spectral performance for the measured data from \autoref{tab:specperf} at the end of the nominal operation phase.} 
\label{tab:specperfdeg}
\begin{center}       
\begin{tabular}{l|r|r|r|r|r}
\multicolumn{2}{c|}{~} & \multicolumn{3}{c|}{LD FWHM} & FD FWHM \\
Emission line & \multicolumn{1}{l|}{Energy}& $t_{exp} = \SI{5.00}{\ms}$ & $t_{exp} = \SI{2.00}{\ms}$ & $t_{exp} = \SI{1.28}{\ms}$ & $t_{exp} = \SI{80}{\us}$ \\
\hline
C K$\alpha_{1,2}$  & \SI{0.2770}{\keV}	& \SI{57.1}{\eV}	& \SI{62.8}{\eV}	& \SI{68.0}{\eV}	& \SI{61.4}{\eV} \\
O K$\alpha_{1,2}$  & \SI{0.5249}{\keV}	& \SI{59.2}{\eV}	& \SI{65.5}{\eV}	& \SI{70.1}{\eV}	& \SI{62.5}{\eV} \\
Zn L$\alpha_{1,2}$ & \SI{1.0117}{\keV}	& \SI{64.2}{\eV}	& \SI{71.1}{\eV}	& \SI{77.4}{\eV}	& \SI{68.3}{\eV} \\
Al K$\alpha_1$     & \SI{1.4867}{\keV}	& \SI{72.7}{\eV}	& \SI{77.9}{\eV}	& \SI{82.1}{\eV}	& \SI{76.5}{\eV} \\
Ag L$\alpha_1$     & \SI{2.9843}{\keV}	& \SI{92.6}{\eV}	& \SI{96.5}{\eV}	& \SI{100.1}{\eV}	& \SI{94.8}{\eV} \\
Ti K$\alpha_1$     & \SI{4.5108}{\keV}	& \SI{113.1}{\eV}	& \SI{119.0}{\eV}	& \SI{121.2}{\eV}	& \SI{115.1}{\eV} \\
Cr K$\alpha_1$     & \SI{5.4147}{\keV}	& \SI{123.0}{\eV}	& \multicolumn{1}{c|}{---}	& \SI{130.3}{\eV}	& \SI{122.4}{\eV} \\
Mn K$\alpha_{1,2}$ & \SI{5.8951}{\keV}	& \SI{127.7}{\eV}	& \SI{131.2}{\eV}	& \SI{133.8}{\eV}	& \SI{130.2}{\eV} \\
Fe K$\beta_{1,3}$  & \SI{7.0580}{\keV}	& \SI{140.0}{\eV}	& \SI{141.8}{\eV}	& \SI{144.3}{\eV}	& \SI{140.4}{\eV} \\
\end{tabular}
\end{center}
\end{table}

\section{Summary}
\label{sec:summary}
For the first time, the two detectors designated for the Wide Field Imager of Athena are under test with their full size as well as their final fabrication technology, pixel layout and readout mode. Beside some noise issues to clarify, they already show an excellent performance over the relevant energy range. A first rough performance analysis indicates, that the energy resolution is degraded only by a few electronvolts until the end of nominal operation. Nevertheless, the spectral resolution is shown only as the mean value of all pixels here. To assess the overall performance, a pixel specific analysis is necessary in the future to quantify the amount of non-compliant pixels.

\acknowledgments 
 
Development and production of the DEPFET sensors for the Athena WFI is performed in a collaboration between MPE and the MPG Semiconductor Laboratory (HLL). We gratefully thank all people who gave aid to make the presented measurements possible. The work was funded by the Max-Planck-Society and the German space agency DLR (FKZ: 50 QR 1901).

\bibliography{report} 
\bibliographystyle{spiebib} 

\end{document}